\documentclass[twocolumn,aps,showpacs,preprintnumbers,amsmath,amssymb]{revtex4}

\usepackage{dcolumn}
\usepackage{graphicx,epsfig}
\usepackage{bm}

\def\be{\begin{equation}}
\def\ee{\end{equation}}

\def\bea{\begin{eqnarray}}
\def\eea{\end{eqnarray}}

\def\Tr{{\rm Tr}\,}

\def\s{\sigma}
\def\a{\alpha}
\def\l{\lambda}
\def\lm{{\l_{\rm max}}}

\begin{document}

\title[Entanglement spectrum in 1D systems]{Entanglement spectrum in one-dimensional systems}

\author{Pasquale Calabrese$^1$ and Alexandre Lefevre$^{2}$}
\affiliation{$^1$Dipartimento di Fisica dell'Universit\`a di Pisa and INFN,
             Pisa, Italy.\\
$^2$IPhT, Orme des Merisiers, CEA Saclay, 91191 Gif sur Yvette Cedex, France.}

\date{\today}

\begin{abstract}
We derive the distribution of eigenvalues of the reduced density matrix of
a block of length $\ell$ in a one-dimensional system in the scaling regime.
The resulting ``entanglement spectrum'' is described by a universal scaling
function depending only on the central charge of the underlying conformal
field theory. This prediction is checked against exact results for the XX
chain. We also show how the entanglement gap closes when $\ell$ is large.
\end{abstract}

\maketitle

The interest in quantifying the entanglement in extended quantum
systems has been growing in recent times at an impressive rate,
mainly because of its ability to detect the scaling behavior in
proximity of quantum critical points (see e.g. Refs. \cite{rev} as
reviews). Among the various measures, the so-called entanglement
entropy has by far been the most studied. By partitioning an
extended quantum system into two blocks, the entanglement entropy is
defined as the von Neumann entropy of the reduced density matrix
$\rho_A$ of one of the two blocks. The success of this quantity can
be understood because it is a single number able to capture the main
features of the scaling behavior. In fact, in one-dimensional (1D)
critical ground-states, when the block $A$ is a segment of length
$\ell$ in an infinite system, the entanglement entropy diverges with
the logarithm of the block size as \cite{Holzhey,Vidal,cc-04} \be
S_A\equiv-\Tr \rho_A\ln \rho_A=\frac{c}3 \ln \ell+ c'_1\,,
\label{SA} \ee where $c$ is the central charge of the associated
conformal field theory (CFT) and $c'_1$ a non-universal constant.
Away from the critical point, $S_A$ saturates to a constant value
\cite{Vidal} proportional to the logarithm of the correlation length
\cite{cc-04}.

However, the reduced density matrix (at least in principle) contains more
information than the entanglement entropy.
This information should be encoded in the full spectrum of the
reduced density matrix, which we shortly call
``entanglement spectrum'', following Ref. \cite{lh-08}.
In this letter we calculate the entanglement spectrum for 1D systems in the
scaling regime, i.e. at or close to a quantum critical point.
The study of the distribution of eigenvalues provides clearly a deeper
theoretical understanding of entanglement and correlations in extended
systems, but not only.
Indeed, all numerical algorithms based on the so-called matrix product
states (among which density matrix renormalization group \cite{dmrg}
is the best known)
give as first output a truncated spectrum of the reduced density matrix
and from this all the other quantities are derived.
Consequently, the knowledge of some scaling properties of this
distribution provides an optimal check for the convergence and the
accuracy of the numerics and could be used for putting accurate bounds on the
efficiency of these methods as already done from other
quantities in Refs. \cite{acc}.

The scaling behavior of the entanglement spectrum can be related to the
properties of the moments of the reduced density matrix:
$R_\a\equiv\Tr \rho_A^\a=\sum_i\l_i^\a$, where $\l_i$ are the eigenvalues of
$\rho_A$.
In the scaling regime, $R_\a$ can be written as
\be
R_\a=c_\a L_{\rm eff}^{-c(\a-1/\a)/6}\,,
\label{Ral}
\ee
where $c_\a$ is a non-universal constant and
$L_{\rm eff}$ is the relevant length in the considered regime.
For example, $L_{\rm eff}$ equals the length block $\ell/a$ if $A$
is part of an infinite gapless system \cite{Holzhey,cc-04},
$L_{\rm eff}=\frac{L}{a}\sin\frac{\pi \ell}L$ if
$A$ is in a finite gapless system of length $L$, and
$L_{\rm eff}=\xi/a$ \cite{cc-04}
if the system is gapped [valid when the correlation
length $\xi$ (the inverse mass gap) is large, but smaller than all the other
lengths like $\ell,L$].
Everywhere $a$ stands for the scale setting the microscopic length, e.g.
the lattice spacing.
The same $\a$ dependence is found in the case of open systems (the exponent of
$L_{\rm eff}$ in Eq. (\ref{Ral}) halves and so does $S_A$ \cite{cc-04})
and when $A$ consists of more disjoint intervals \cite{cc-04} and also in some
non-equilibrium situations \cite{neq}.
For practical reasons, it is convenient to write
\be
R_\a=c_\a e^{-b (\a-1/\a)},\mbox{ with }\;
b=\frac{c}6 \ln L_{\rm eff}>0\,,
\label{Ra}
\ee
where by simple inspection $b$ is related to the maximum
eigenvalue: $b=-\ln \lm$, as well known \cite{singcopy}.

In order to characterize the entanglement spectrum, we define the
{\it distribution of eigenvalues} $P(\l)=\sum_i \delta(\l-\l_i)$,
which is normalized to $m^\ell$, where $m$ is the dimension of the
local Hilbert space, e.g. $m=2$ for spin $1/2$ systems. We determine
this distribution for a 1D system in the scaling regime where Eq.
(\ref{Ral}) applies. The only assumption in what follows is that the
$\alpha$-dependence of the non-universal part of $c_\a$ can be
ignored, i.e. that we can write $c_\a= a^{c(\a-1/\a)/6} f_\a$, with
$f_\a$ a constant function (i.e. the effect of $c_\a$ is to replace
the lengths appearing in $R_\a$ with the dimensionless quantity
$L_{\rm eff}$). A priori this can appear as a very crude
approximation, but it is not the case. For the gapless XX chain,
$c_\a$ is known analytically \cite{jk-04}, and it is easy to show
that $f_\a$ varies less then 1\% as soon as $\a\geq 2$. Physically,
this is equivalent to stating that the main contribution of $c_\a$
is to set the microscopic length $a$ to be used in the continuum
description in terms of the lattice one. We will check a posteriori
for the XX chain that the result obtained ignoring $c_\a$ describes
accurately the actual entanglement spectrum when $L_{\rm eff}$ is
large.
A very precise numerical determination of $c_\a$ for the XXZ
chain with $-1<\Delta\leq1$ \cite{ccn-08} shows that the same scenario is true
on the full critical line of the model.
In few other cases $c_\a$ is also known \cite{ch-05,ccd-07,fik-08} and similar
properties can be found.

Thus, ignoring temporarily $c_\a$, we will compute the entanglement
spectrum, using the simple observation that
$\l P(\l)=\lim\limits_{\epsilon\rightarrow 0} {\rm Im}\,f(\l-i\epsilon)$, with
\be
 f(z)=\frac{1}{\pi}\sum_{n=1}^\infty R_n z^{-n}=\frac{1}{\pi}\int
 d\l\frac{\l P(\l)}{z-\l}.
\label{fz}
\ee
Here, $f(\l-i\epsilon)$ has an imaginary part when  $\epsilon\rightarrow 0$ only on the support of $P(\lambda)$, due to the pole in the r.h.s. of (\ref{fz}).
The calculation of $f(z)$ is straightforward:
\be
f(z)=
\frac1\pi \sum_{k=0}^\infty \frac{b^k}{k!}
\sum_{n=1}^\infty \frac{(e^{-b}/z)^n}{n^k}=
\frac1\pi \sum_{k=0}^\infty \frac{b^k}{k!} {\rm Li}_k(e^{-b}/z),
\label{li}
\ee
where ${\rm Li}_k(y)$ is the polylogarithm function, which is analytic on the
complex plane, with a cut on the real axis for $y\geq 1$ (that once
again is just $b=-\ln \lm$).
The discontinuity along the cut is given by
$\displaystyle\lim_{\epsilon\to0} {\rm Im}\, {\rm Li}_k(y+i\epsilon)=
\pi (\ln y)^{k-1}/\Gamma(k)$ for $k\geq 1$.
The sum can be explictly done and we end up with
\bea
P(\l)&=&\delta(\l_{\rm max}-\l)+\nonumber
\\&+&
\frac{b\,\theta(\lm-\l)}{\l\sqrt{b\ln(\l_{\rm max}/\l)}}
I_1(2\sqrt{b \ln(\l_{\rm max}/\l)}),\label{PL}
\eea
where the delta peak comes from the contribution of $k=0$ in (\ref{li}) and
$I_k(x)$ stands for the modified Bessel function of the first kind.
In this derivation, only $R_\a$ with positive integers $\a$ enter.
For these values (as we already discussed) $c_\a$ in general
does not vary significantly.
This gives an argument that explains why Eq. (\ref{PL}) works well
for large enough $L_{\rm eff}$ in the following calculations for the XX model
and why the same is expected in general.

$P(e^{-t})$ can also be obtained by considering the inverse Laplace
transform of $R_\a$ in the variable $\a$. Using standard results
Eq. (\ref{PL}) is easily recovered.
We preferred to give the previous derivation because it highlights the role
played by positive integer $\a$, where $c_\a$ can be ignored.
Conversely the inverse Laplace transform requires an integral on
the complex plane, over a contour where there is less control on the values of
$\a$ that are contributing relevantly.
We noticed this because numerical inverse Laplace transform could be
done in principle in more complicated cases (like those in
Refs. \cite{jk-04,ccn-08,ch-05,ccd-07,fik-08})
where the previous analytic reasoning does not work.

Let us discuss now the main properties of $P(\l)$:

(i) {\it The mean number of eigenvalues} larger than a given $\l$ is
\be
n(\l)=\int_\l^{\l_{\rm max}} d\l P(\l) =
I_0(2\sqrt{b \ln (\lm/\l)})\,.
\label{nlam}
\ee
Note that for $\l\to 0$, $n(\l)$ diverges, as it should,  because in the
continuum the number of eigenvalues is infinite. In the lattice models,
this can be regularized by the finite number of degrees of freedom (as e.g.
in spin chains), but not always (e.g. bosons have always infinitely many
eigenvalues).

(ii){\it The normalization} $\sum \l_i=1$ corresponds to $\int \l P(\l)=1$,
and follows directly from Eq. (\ref{PL}).


(iii) {\it The entanglement entropy} is given by
\be
S= -\int_0^\lm \l \ln\l P(\l) d\l=-2\ln \lm\,, \ee reproducing the
result that the single copy entanglement equals one-half of the
entanglement entropy \cite{singcopy}.

(iv) {\it Majorization} is a relation between two probability
distributions $\l\equiv\{\l_i\}$ and $\mu\equiv\{\mu_i\}$ whose
elements are ordered $\l_1>\l_2\dots >\l_N$ (and similarly for
$\mu$): it is said that $\l$ majorizes $\mu$ if $\sum_{i=1}^M \l_i
\geq\sum_{i=1}^M \mu_i $ for any $M=1,\dots,N$ and $\sum_{i=1}^N
\l_i =\sum_{i=1}^N \mu_i=1$. It has been argued, observed
numerically and in some instances proven analytically, that with
increasing $L_{\rm eff}$ the resulting distribution of eigenvalues
is majorized by the ones at smaller scaling lengths
\cite{maj,zbfs-06} (sometimes this is referred to as majorization
along renormalization group flow). From the previous result it is
straightforward that majorization holds in the scaling regime when
Eq. (\ref{PL}) for $P(\l)$ applies. In fact, we have \be s(M)\equiv
\sum_{i=1}^M \l_i \to \lm\left[1+\int_0^{I_0^{-1}(M)} dy e^{-y^2/4b}
I_1(y)\right], \label{srformula} \ee which, at fixed $M$, is a
monotonous function of $\lm$ (that is a monotonous function of
$L_{\rm eff}$). This proves majorization in a very easy way. It is
also simple to check majorization directly by making the numerical
integral at fixed $M$ and varying $\lm$.

\begin{figure}
\includegraphics[width=.47\textwidth, clip]{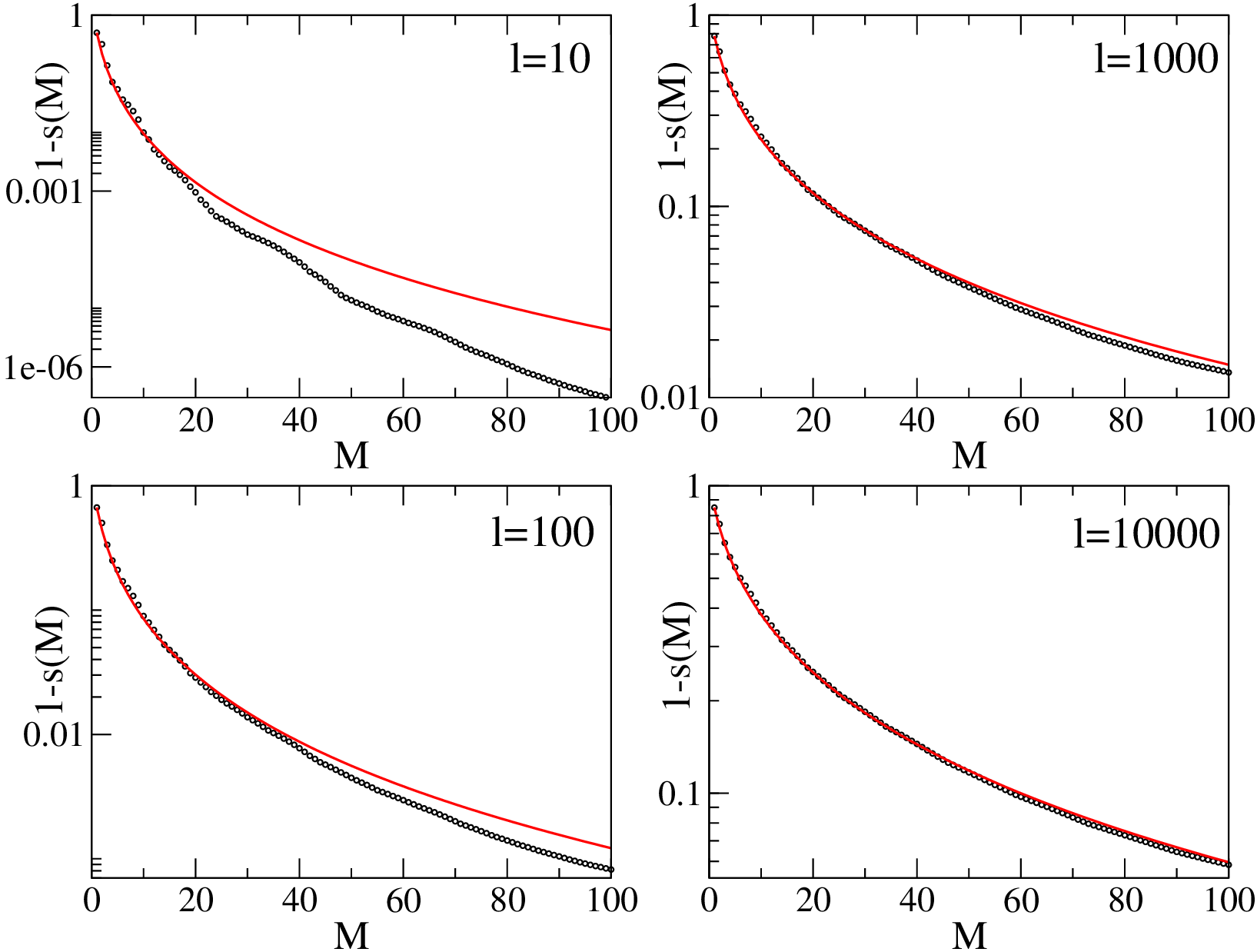}
\caption{Sum of the first $M$ eigenvalues of the XX model up to $M=100$:
$1-s(M)$ as function of $M$ for $\ell=10,100,1000,10000$ (black dots).
The red line is the conformal field theory prediction Eq. (\ref{srformula}),
in which $\lm$ has been fixed to the maximum eigenvalue obtained numerically.
}
\label{sumrule}
\end{figure}

We now compare the scaling function for the entanglement spectrum
with the eigenvalues of lattice models, in order
to show its predictivity and to highlight its limits.
As a prototype of lattice models we consider the gapless XX chain defined
by the Hamiltonian
\be
H_{XX}=\sum_i[\s_i^x\s_{i+1}^x+\s_i^y\s_{i+1}^y]\,,
\ee
where $\s_i^{x,y}$ are the Pauli matrices at site $i$.
The reduced density matrix of a block of $\ell$ contiguous spins
in an infinite chain (so, among the cases of before, this corresponds
to $L_{\rm eff}=\ell/a$) 
can be obtained by exploiting the mapping to free fermions \cite{pesc,Vidal}:
\be
\rho_\ell\propto \exp\left[-\sum_{1\leq i,j\leq \ell} h_{ij} c^\dagger_i
  c_j\right],
\ee
where $c_i$ are spinless fermion annihilation operators, and
\be
h=\ln [(1-C)/C]\,, \qquad C_{ij}=\frac{\sin \pi (i-j)/2}{\pi (i-j)}\,.
\ee

\begin{figure}[b]
\includegraphics[width=0.43\textwidth,clip]{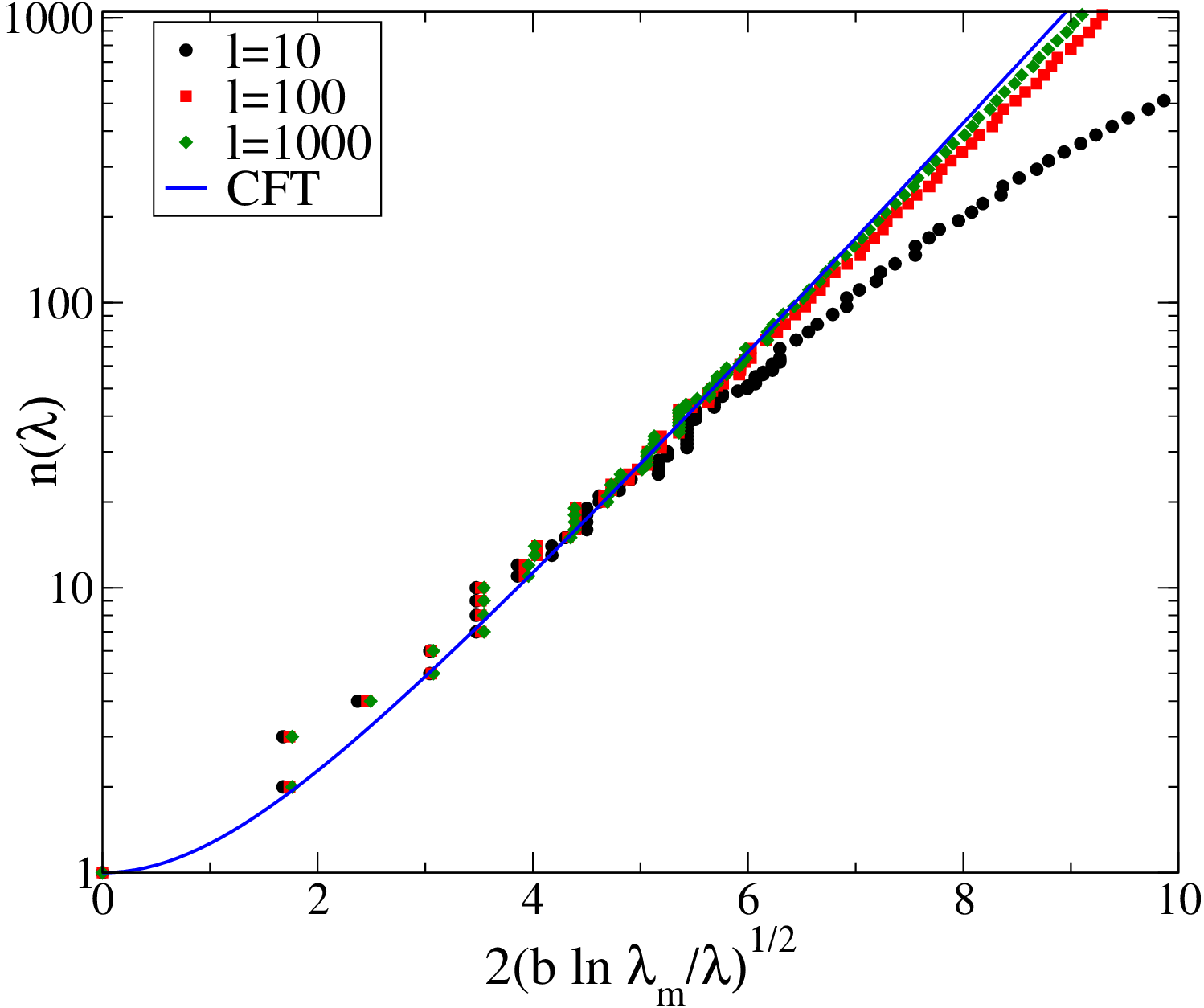}
\caption{Inverse function of $\l_i$ (value of the $i$-th eigenvalue of the
  reduced density matrix) for $\ell=10,100,1000$ for the XX chain.
  The plot is shown in terms of the scaling variable $2\sqrt{b\ln (\lm/\l)}$.
  The full line is the CFT prediction.
}
\label{rhoeigcut}
\end{figure}

Calculating the eigenvalues $\nu_i$ of the matrix $C_{ij}$ requires
only the diagonalization of an $\ell\times\ell$ matrix. In terms of
the $\nu_i$, the $2^\ell$ eigenvalues of the reduced density matrix
are the products \be \l_{A} =\prod_{i\in A, j\in B} \nu_i
(1-\nu_j)\,, \ee where $A$ is a subset of $\{1,2,\dots, \ell\}$ and
$B$ the complement. While it is possible to obtain all the $\nu_i$
up to very large $\ell$, making all of the $2^\ell$ products to
obtain the full spectrum requires far too much memory on a personal
computer. Thus we calculate the full eigenvalue spectrum only up to
$\ell=28$ (which corresponds to almost $300$ millions of
eigenvalues), and for larger $\ell$ we truncated the fermionic
spectrum to $24$ modes (i.e. we diagonalize the $C$ matrix finding
all the $\nu_i$, but in the products we do not vary the $\ell-24$
eigenvalues that are closer to $0$ or $1$ and generate the smaller
$\l_i$). This reproduces in the exact order the first few thousands
eigenvalues, to which we limit our analysis in the truncated cases.

We start this analysis from the check of the
``sumrule'' given by Eq. (\ref{srformula}).
In Fig. \ref{sumrule} we report $1$ minus the sum
of the first $M$  (up to $100$) eigenvalues for $\ell=10,100,1000,10000$.
It is evident that increasing $\ell$ the sum is well described by
Eq. (\ref{srformula}) for larger and larger values of $M$.
The (negative) deviations from the CFT take place at a value of $M$ that
roughly scales like $\ln\ell$ and they are obviously due to lattice effects,
because the sum rule must be saturated by a finite sum up to $2^\ell$ that
cutoffs the integral of the continuum limit.

\begin{figure}[t]
\includegraphics[width=.48\textwidth,clip]{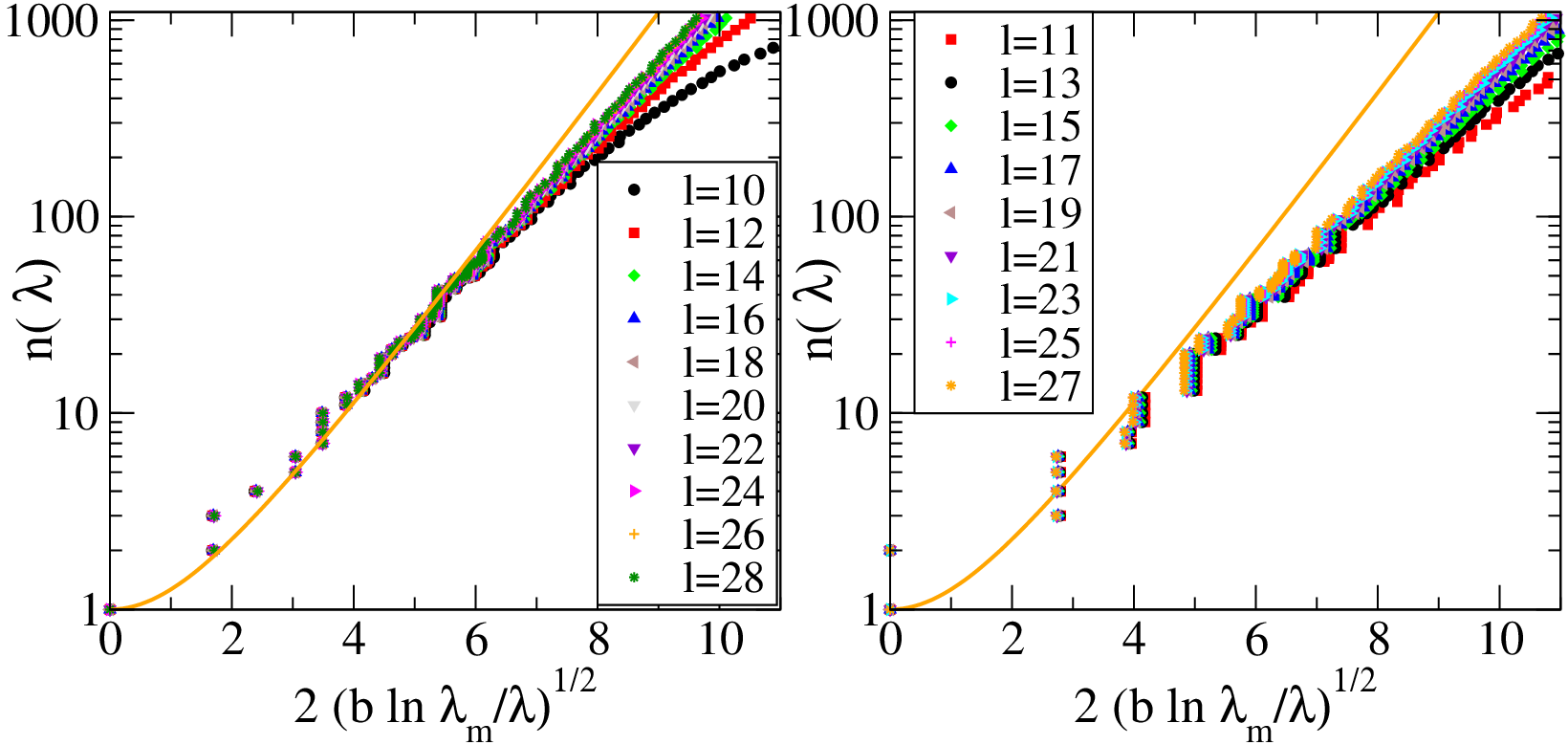}
\caption{The same as in Fig. \ref{rhoeigcut}  for all $10\leq\ell\leq28$.
Even (odd) $\ell$ are shown on the left (right) panel.
}
\label{rhoeigex}
\end{figure}

A stronger test of our predictions is provided by Eq. (\ref{nlam}) that gives
the total number of eigenvalues between $\l$ and $\lm$.
This formula provides the natural scaling variable
$x=2\sqrt{b\ln (\lm/\l)}$.
Once $\lm$ is determined from the numerics, $n(\l)$ is just given by $I_0(x)$,
independently from any other detail of the model under consideration and,
more surprisingly, it is also independent from the model itself.
Instead of plotting the number of eigenvalues larger than a given $\l$,
we prefer to show the inverse function of $\l_i$ that is the value of the
$i$-th eigenvalue.
In addition to represent $n(\l)$, this also has the advantage of giving
information
about the degeneracy of all eigenvalues, as the
number of points at the same position on the horizontal axis.
For $\ell=10,100,1000$ this function is shown in Fig. \ref{rhoeigcut} and
compared with $n(\l)$ in Eq. (\ref{nlam}).
It is evident that when the spectrum becomes almost continuous, it is well
described by the CFT prediction up to a given $\l_{\rm min}(\ell)$ (i.e.
a given $x_{\rm  max} (\ell)$).
Increasing $\ell$ the number of eigenvalues described by Eq. (\ref{nlam})
increases and at $\ell=1000$, approximately thousands eigenvalues fall on
the CFT scaling curve.
The negative deviations for eigenvalues smaller than $\l_{\rm min}(\ell)$
are due to
lattice effects as in the previous case of the partial sumrule.
The degeneracies of the eigenvalues are not reproduced by our approach.
In Fig. \ref{rhoeigex} we plot the entanglement spectrum for all $\ell$
from $10$ to $28$ in order to show both finite-size and parity effects.
When $\ell$ is odd (a case rarely considered in the literature)
the approach to the asymptotic scaling is much slower.
This can be understood because of the double degeneracy
of {\it all} the eigenvalues (included the largest), which moves a large
weight toward $x=0$ that, because of the sumrule, must be compensated
by a smaller weight at large $x$ (i.e. small eigenvalues).

There is a final interesting feature that is clearly visible in
Figs. \ref{rhoeigcut} and \ref{rhoeigex}.
When plotting in terms of $x=2\sqrt{b\ln (\lm/\l)}$, the first few
eigenvalues do not change their positions with changing $\ell$.
The logarithms of these discrete eigenvalues result to be equispaced, i.e.
\be b\ln \frac{\l_\mu}{\l_\nu}\simeq k \Rightarrow\quad
\frac{\l_\nu}{\l_\mu}\simeq e^{-\frac{6 k}{\ln\ell/a}}, \label{gaps}
\ee where $\mu$ and $\nu$ are indices referring to two
non-degenerate consecutive eigenvalues and  $k$ is a constant that
is different for even and odd $\ell$ (and we used $b=-\ln \lm=c/6
\ln\ell$ and $c=1$). When $\l_\mu=\lm$, this simple formula tells us
how the gap between the first and the second eigenvalues closes when
increasing $\ell$, which, following Ref. \cite{lh-08}, we call
"entanglement gap". This scaling of the discrete part of the
spectrum can be related to an old result about the scaling of the
eigenvalues of the corner transfer matrix \cite{pt-87}. In fact,
because $b\ln (\l_\mu/\l_\nu)$ is almost independent of $\ell$, for
this part of the spectrum the result should be the same as that of a
segment $\ell$ in a system of length $2\ell$. In this case the
reduced transfer matrix can be seen as the fourth power of a corner
transfer matrix \cite{cc-04} (of angle $\pi/2$), for which the
analogous of Eq. (\ref{gaps}) has been derived from CFT
\cite{pt-87}. Using this correspondence the asymptotic behavior of
$P(\l\ll 1)$ for {\it gapped} systems has been already derived in
Ref. \cite{jap}. However our result goes far beyond, explaining also
the reason of such a simple result.


In conclusions, the main result of this letter is the analytic
derivation of the {\it universal} entanglement spectrum for 1D
models in the scaling regime, given by Eq. (\ref{PL}). This turns
out to depend only on the central charge of the underlying conformal
field theory. We found that Eq. (\ref{PL}) describes accurately the
continuum part of the spectrum  of the XX chain for large enough
$L_{\rm eff}$, and is expected to work for any model at or close to
a quantum critical point. This can be seen as a surprise because it
means that the continuum part of the entanglement spectrum does not
contain more information than the entanglement entropy that is just
one possible average over the spectrum. The reason for this (maybe
unexpected) result can be traced back to the fact that conformal
invariance in 1D is so strong that it fixes completely the shape of
the full spectrum, leaving only one parameter (the central charge)
free. Such parameter can be fixed by of a single ``measure'' like
the entanglement entropy, the largest eigenvalue, etc.

Oppositely, the discrete part of the spectrum is only
reproduced in average by Eq. (\ref{PL}). As a consequence, the location and
the degeneracies of the low-lying eigenvalues of the reduced density matrix
can still be a tool for extracting further universal information on the model
under investigation that could not be captured by the entanglement entropy.
For example, it is known that the degeneracies of eigenvalues of
the isotropic Heisenberg antiferromagnetic chain are larger than the XXZ ones
(because of the larger symmetry), but they all have $c=1$ and so the same
continuum part of the spectrum.
A careful study of these issues is needed,
but it goes far beyond the goals of this letter.

{\it Acknowledgments}.
This work started from a discussion with J. Cardy.
We thank L. Tagliacozzo for pointing out the important Ref. \cite{pt-87}.
We thank J. Cardy, A. Celi, S. Iblisdir, J. I. Latorre, B. Nienhuis,
and L. Tagliacozzo for stimulating discussions.
PC benefited of a travel grant from ESF (INSTANS activity).


\end{document}